\documentclass[journal]{IEEEtran}
\IEEEoverridecommandlockouts
\usepackage{cite}
\usepackage{amsmath,amssymb,amsfonts}
\usepackage{algorithmic}
\usepackage{graphicx}
\usepackage{textcomp}
\usepackage[dvipsnames]{xcolor}
\usepackage{url}

\usepackage{selinput}\SelectInputMappings{adieresis={ä},germandbls={ß}}
\usepackage{algorithmic}
\usepackage[ruled,vlined,linesnumbered,commentsnumbered]{algorithm2e}
\usepackage{leftidx}
\usepackage{amsthm}
\usepackage[capitalise]{cleveref}
\usepackage{soul}
\usepackage{svg}
\usepackage{subfigure}
\usepackage{array}
\usepackage{multirow}
\usepackage{hhline}
\usepackage[font=footnotesize]{caption}
\usepackage{booktabs}
\usepackage{stfloats}
\newcolumntype{L}[1]{>{\raggedright\let\newline\\\arraybackslash\hspace{0pt}}m{#1}}
\newcolumntype{C}[1]{>{\centering\let\newline\\\arraybackslash\hspace{0pt}}m{#1}}
\newcolumntype{R}[1]{>{\raggedleft\let\newline\\\arraybackslash\hspace{0pt}}m{#1}}

\def\BibTeX{{\rm B\kern-.05em{\sc i\kern-.025em b}\kern-.08em
    T\kern-.1667em\lower.7ex\hbox{E}\kern-.125emX}}

\usepackage[shortcuts,acronym]{glossaries}
\newacronym{auc}{AUC}{Area Under the receiver operating characteristic Curve}
\newacronym{aeica}{AeICA}{Adaptive extraction-based ICA}
\newacronym{ai}{AI}{artificial intelligence}

\newacronym{bss}{BSS}{Blind Source Separation}

\newacronym{cdf}{CDF}{Cumulative Distribution Function}
\newacronym{cots}{COTS}{Commercial off-the-shelf}

\newacronym{coin}{COIN}{Computing in the Network}
\newacronym{comnetsemu}{ComNetsEmu}{Communication Networks Emulator}
\newacronym{cs}{CS}{Compressed Sensing}
\newacronym{cf}{C-F}{Compute-and-Forward}
\newacronym{cp}{CP}{control plane}
\newacronym{cups}{CUPS}{control and user plane separation}
\newacronym{cnn}{CNN}{convolutional neural network}

\newacronym{de}{DE}{Data Extraction}
\newacronym{dl}{DL}{deep learning}
\newacronym{dn}{DN}{data network}
\newacronym{dnn}{DNN}{deep neural network}
\newacronym{dp}{DP}{data plane}
\newacronym{dr}{DR}{Data Reconstruction}

\newacronym{e2e}{E2E}{end-to-end}
\newacronym{easdf}{EASDF}{Edge Application Service Discovery Function}

\newacronym{ica}{ICA}{Independent Component Analysis}
\newacronym{iot}{IoT}{Internet of Things}
\newacronym{iiot}{IIoT}{Industrial Internet of Things}

\newacronym{mimii}{MIMII}{Malfunctioning Industrial Machine Investigation and Inspection}
\newacronym{mtu}{MTU}{Maximum Transmission Unit}
\newacronym{mauc}{mAUC}{mean-AUC}
\newacronym{ml}{ML}{machine learning}
\newacronym{mp}{MP}{management plane}
\newacronym{mano}{MANO}{Management and orchestration}

\newacronym{nf}{NF}{Network Function}
\newacronym{nfv}{NFV}{Network Function Virtualization}
\newacronym{nn}{NN}{neural network}

\newacronym{pauc}{pAUC}{partial-AUC}
\newacronym{pica}{pICA}{progressive ICA}
\newacronym{pisa}{PISA}{Protocol-Independent Switch Architecture}
\newacronym{pdu}{PDU}{packet data unit}

\newacronym{qos}{QoS}{Quality-of-Service}

\newacronym{roc}{ROC}{Receiver Operating Characteristic}
\newacronym{rrt}{RTT}{Round Trip Time}

\newacronym{sdn}{SDN}{Software Defined Networking}
\newacronym{sfc}{SFC}{Service Function Chain}
\newacronym{sme}{SME}{Separation Matrix Estimation}
\newacronym{smf}{SMF}{Session Management Function}
\newacronym{sf}{S-F}{Store-and-Forward}
\newacronym{svm}{SVM}{Support-Vector Machine}
\newacronym{sdr}{SDR}{Source-to-Distortion Ratio}

\newacronym{tcp}{TCP}{Transmission Control Protocol}

\newacronym{ue}{UE}{user entity}
\newacronym{urllc}{URLLC}{Ultra-Reliable and Low-Latency Communications}
\newacronym{udp}{UDP}{User Datagram Protocol}
\newacronym{upf}{UPF}{user plane function}

\newacronym{vnf}{VNF}{Virtualized Network Function}

\newacronym{vnfm}{VNFM}{Virtualized Network Function Manager}

\newacronym{vim}{VIM}{Virtualized Infrastructure Manager}

\newacronym{rtt}{RTT}{Round-Trip-Time}

\newcommand{\rv}[1]{\textcolor{black}{#1}}

\begin{document}
\bstctlcite{IEEEexample:BSTcontrol}
\title{Functional Split of In-Network Deep Learning for 6G: A Feasibility Study
\thanks{\copyright \; 2022 IEEE. Personal use of this material is permitted. Permission from IEEE must be obtained for all other uses, in any current or future media, including reprinting/republishing this material for advertising or promotional purposes, creating new collective works, for resale or redistribution to servers or lists, or reuse of any copyrighted component of this work in other works.}
}
\author{
    \IEEEauthorblockN{
        Jia He\IEEEauthorrefmark{1},
        Huanzhuo Wu\IEEEauthorrefmark{1},
        Xun Xiao\IEEEauthorrefmark{3},
        Riccardo Bassoli\IEEEauthorrefmark{1}\IEEEauthorrefmark{2}\IEEEauthorrefmark{4},
        and
        Frank H. P. Fitzek\IEEEauthorrefmark{1}\IEEEauthorrefmark{4}\\
    }
    \IEEEauthorblockA{\IEEEauthorrefmark{1}Deutsche Telekom Chair of Communication Networks - Technische Universität Dresden, Germany
    }\\
    \IEEEauthorblockA{\IEEEauthorrefmark{2}\rv{Quantum Communication Networks Research Group - Technische Universität Dresden, Germany
    }}\\
    \IEEEauthorblockA{\IEEEauthorrefmark{3}Munich Research Center, Huawei Technologies, Muenchen, Germany}\\
    \IEEEauthorblockA{\IEEEauthorrefmark{4}Centre for Tactile Internet with Human-in-the-Loop (CeTI)}
}

\markboth{ACCEPTED FOR PUBLICATION IN THE IEEE WIRELESS COMMUNICATIONS MAGAZINE, DOI: 10.1109/MWC.003.2200060}{}

\maketitle

\begin{abstract}
In existing mobile network systems, the data plane (DP) is mainly considered a pipeline consisting of network elements end-to-end forwarding user data traffics. With the rapid maturity of programmable network devices, however, mobile network infrastructure mutates towards a programmable computing platform. Therefore, such a programmable DP can provide in-network computing capability for many application services. In this paper, we target to enhance the data plane with in-network deep learning (DL) capability. However, in-network intelligence can be a significant load for network devices. Then, the paradigm of the functional split is applied so that the deep neural network (DNN) is decomposed into sub-elements of the data plane for making machine learning inference jobs more efficient. As a proof-of-concept, we take a Blind Source Separation (BSS) problem as an example to exhibit the benefits of such an approach. We implement the proposed enhancement in a full-stack emulator and we provide a quantitative evaluation with professional datasets. As an initial trial, our study provides insightful guidelines for the design of the future mobile network system, employing in-network intelligence (e.g., 6G).

\end{abstract}

\begin{IEEEkeywords}
Artificial Intelligence, In-Network Computing, Blind Source Separation, Data Plane Enhancement, 6G
\end{IEEEkeywords}

\section{Introduction}\label{sec:introduction}

With the advent of network softwarization and 5G, the mobile network system consists of a \ac{cp} and a \ac{dp}. \ac{cp} serves the purposes of access control, security, handling service requests, policy/charging, etc.; \ac{dp} consists of network resource elements to deploy network services commanded by \ac{cp}. Before 5G, \ac{cp} and \ac{dp} were tightly coupled, in vendor-specific hardware devices. 

5G decoupling allows \ac{cp} and \ac{dp} to evolve separately and flexibly \rv{by} \ac{nfv} and \ac{sdn} enablers of network softwarization. 
Currently, both \ac{cp} and \ac{dp} \acp{nf} can be fully virtualized, interconnected as software components in virtual chains, deployed wherever network resources are available.


Recent development in programmability has opened the way for in-network intelligence, which is going to be a key aspect of future 6G networks.
P4 language is used in softwarization and its performance is \rv{nearly} equivalent to those traditional vendor-specific products. Examples can be found in model training~\cite{Lao2021ATPnetworkAggregation}, distributed database consensus~\cite{Dang2020P4xosConsensusas}, etc. 
\rv{These pioneer works show a chance to bring application services back to operator's networks if the \ac{dp} becomes fully programmable in the forthcoming 6G.}

In this paper, driven by the key importance of \ac{ai} in various vertical industries, we consider a \ac{dp} enhancement with in-network \ac{dl} capability. Specifically, we extend configurable \acp{upf} with a shredding \ac{dnn} after modeling training to support in-transit inference services over the extended \acp{upf} deployed in the \ac{dp}. In order to exemplify the benefits of such an enhancement in \ac{dp}, we take \ac{bss} problem as a case study. A \ac{bss} problem is a mixture data separation problem having many realistic applications e.g., audio analysis, natural language processing, speech recognition, etc. It is representative of a set of typical inference problems that can be solved by \ac{dnn}-based approaches. We expect that these will occupy the majority of future applications in 6G. 

The unique features of our work are as follows. First, we consider \ac{dp} for \ac{dnn} inference tasks, complementary to model training problems widely studied in the literature~\cite{Lao2021ATPnetworkAggregation}. Secondly, our problem is a special type of distributed \ac{ai} problems as the inference job has to be done in sequential order over the extended \acp{upf} of a forwarding path in \ac{dp}. In other words, our problem has to take the order of \acp{upf} into account when designing a \ac{dnn} split-and-deployment strategy. In contrast, usual distributed learning problems do not have such a constraint. Last but not least, the \ac{dp} enhancement also requires modifying the counterpart \ac{cp} procedures/interfaces. In summary, our contributions are briefly listed as follows:
\begin{itemize}
	\item We enhanced \ac{dp} with in-network \ac{dl} capability in order to support popular \ac{dnn} inference application services for end-users. Specifically, We proposed \ac{dnn} split strategies in order to realize a progressive data processing for \ac{ai} inference tasks; accordingly, extended \ac{cp} interfaces and procedures are introduced as well;
	\item We took a \ac{bss} problem as an example to demonstrate how \acp{upf} can be enhanced with accommodating split \ac{dnn}. Particularly, we converted a monolithic \ac{cnn} Conv-TasNet~\cite{Luo2019ConvtasnetSurpassing} into a split version that can be deployed on the extended \acp{upf};
	\item We implemented our proposed solutions in a full-stack emulator--\ac{comnetsemu}. We compared with its original monolithic version and the latest non-\ac{dl}-based solution~\cite{picaextention}, the evaluation results based on a professional dataset confirm the benefits of the \ac{dp} intelligence enhancement.
\end{itemize}
To the best of our knowledge, we are not aware of similar works in the literature. In addition, our work well aligns with 3GPP prospective work items for the upcoming 5G Release 18, targeting to support \ac{ai} applications, and the design of future 6G networks. Our work also provides insightful guidelines on the impact of \ac{ai} on the design of open programmable networks in 6G. 

The rest of the paper is organized as follows.
First, the related works on in-network \ac{dl} are introduced in~\cref{sec:related_work}. 
\cref{sec:data_plane_enhancement} describes the general strategy for \ac{dl} enhancement. After that, a \ac{bss} problem is used to showcase the proposed scheme in~\cref{sec:bss_example}.
\cref{sec:evaluation} covers the full-stack implementation, the quantitative evaluation, and in-depth discussions about our measurement results.
Finally, \cref{sec:conclusion} concludes our contributions and points out future research aspects.
\section{Related work}\label{sec:related_work}
\rv{This section introduces the related works on programmable network devices and in-network \ac{dl}.}

\subsection{Programmable network devices}
Network softwarization technologies decouple the \ac{cp} and \ac{dp} structure of the network to make in-network computing possible.
One of the main advantages of this is that the network endpoints (e.g., servers) suffer from reduced computing power as the workload is offloaded to the network~\cite{Wu2021ComputingMeetsNetwork}.

The drive for programmability of network devices has a long history, such as \ac{pisa}, smartNIC, etc.
With the advance of \ac{sdn}, \acp{vnf} are interconnected through the underlying infrastructure network as \rv{containers, virtual machines}, or natively as bare-metal.
This allows the creation of an \ac{sfc} that performs the required computation on the physical devices.

On this basis, P4 -- a high-level language to configure programmable network devices -- was proposed.
The P4 compiler translates P4 programs into code that runs on the underlying switches. Thus, P4 serves as a common interface between \ac{sdn} controllers and programmable network devices~\cite{Dang2020P4xosConsensusas}.
These trends stimulate the concept of in-network \ac{dl}, which naturally fits into an edge-cloud continuum within a common framework integrated compute-network capabilities.

\subsection{In-network \rv{\ac{dl}}}
Within the broad area of in-network \ac{dl}, there have been several streams of research. One stream of research focuses on enabling \ac{dnn} training by using distributed learning.
\cite{Lao2021ATPnetworkAggregation} supports in-network aggregation to enable sharing switch resources across training jobs.
\cite{Xu2021GRACECompressedCommunication} considers the overhead issues of distributed learning problems.
A unified traffic compression framework for distributed learning is proposed, which can be easily implemented with most toolkits.
\rv{In~\cite{DBLP:journals/corr/abs-2107-03428}, authors explored the challenges and future directions for federate learning on edge devices.}
In~\cite{Liang2020edgebaseddeep}, an edge computing-based system is designed to offload the data analysis process from the cloud by deploying the same \ac{cnn} model to edge nodes and feeding them with different data subsets.
These efforts deploy the deep learning network as a whole on a network device, while the end server is mainly responsible for parameter synchronization and updating during the training phase.
For training models, these approaches have obvious advantages, such as distributing the training tasks and reducing the data communication volume. However, these advantages cannot be retained in the inference phase, where parameter learning is no longer required.

Another stream has focused on deploying and running \ac{dnn} in the \ac{dp}.
The authors of~\cite{Xavier2021ProgrammableSwitchesNetworking} transformed the machine learning model for network traffic classification into a P4 language pipeline and deployed it on programmable switches via an Agent Deployer. However, this study \rv{focused} on the design of the \rv{\ac{nn}} and the interface between \ac{cp} and \ac{dp}. Splitting \rv{\acp{nn}} into multiple \acp{vnf} to be deployed is not covered.
\cite{Saquetti2021NetworkIntelligenceRunning} provides the first open-source implementation of a distributed \rv{\ac{nn}} in a programmable \ac{dp}.
This paper is based on the deployment of neurons, however, as \acp{dnn} grow increasingly large, the number of neurons explodes much larger than the available network devices, making neuron-based deployment in the network impractical.

While these streams have made important and unique contributions to in-network \ac{dl}, the need for a splitting strategy to distribute \acp{dnn} is becoming more and more intense, especially with today's programmable network devices and increasingly large-scale \rv{\ac{nn}} models.
\section{In-network \ac{dl} Enhancement for 6G}\label{sec:data_plane_enhancement}
\rv{In this section, our proposed extension on \ac{dp} is presented, followed by an introduction to the \ac{mp} and \ac{cp} extension for 6G.}

\begin{figure}[t!]
	\centerline{\includegraphics[width=\columnwidth]{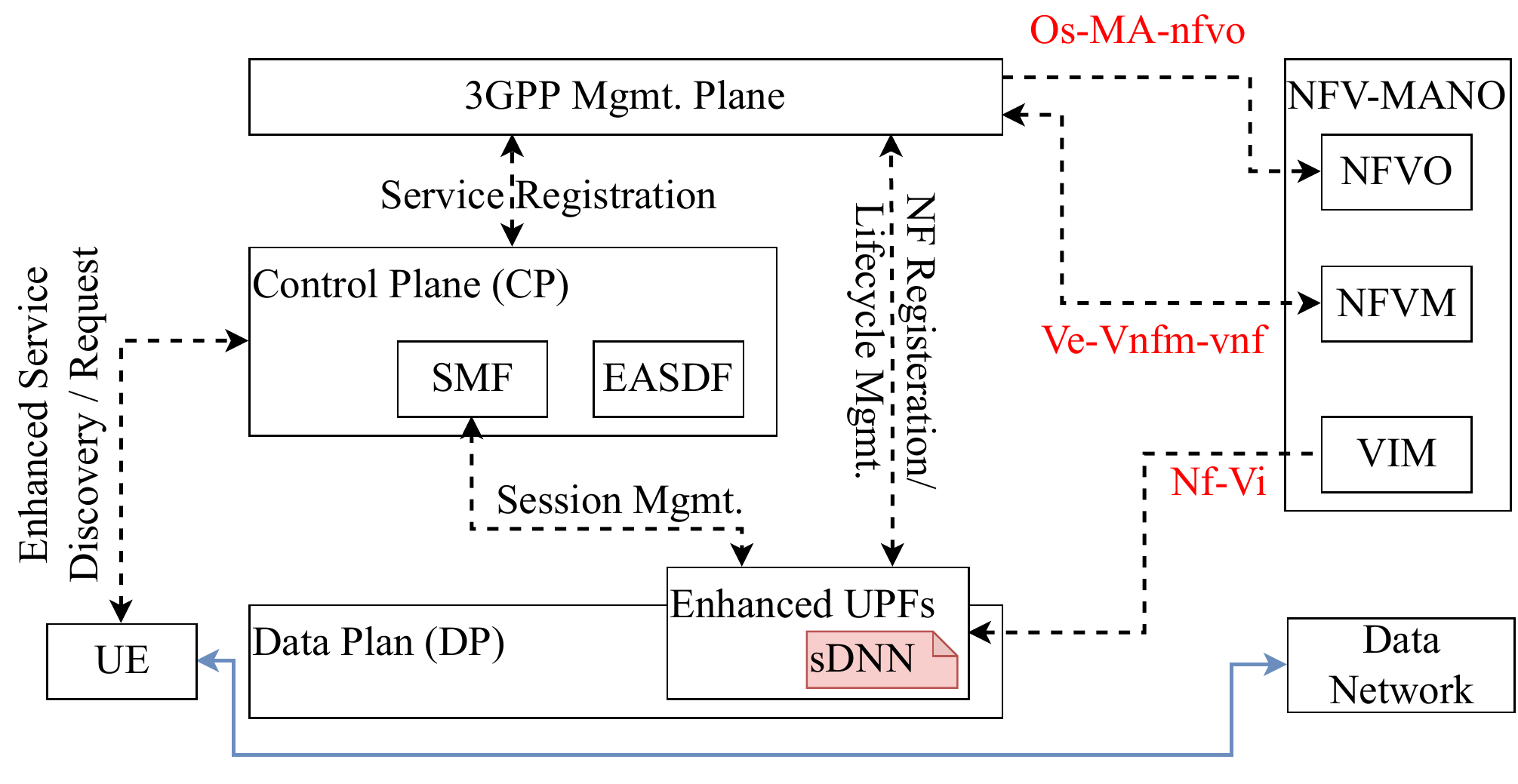}}
	\caption{\ac{dl} capability enhancements in a 3GPP system. \rv{For more information on the interface between NFV-MANO and MP/DP, please refer to the NFV-IFA group specification published by the European Telecommunications Standards Institute.}}
	\label{fig:sysenhance}
\end{figure}

\subsection{\rv{\ac{dp}} \ac{dl} capability extension}\label{subsec:dpdl}
According to the reference architecture defined by 3GPP SA2, in core network part, \ac{dp} mainly consists of \acp{upf} transporting \ac{pdu} sessions. A \ac{pdu} session represents \rv{a} QoS-aware connection between a \ac{ue} and a \rv{data network} located at the edge or in a data center. As mentioned, in existing mobile network systems, \ac{upf} acts as a pure networking element, only transmitting bits and bytes in terms of the IP header of a data packet. In order to extend \ac{upf} with \ac{dl} capability, it is equivalent to enhancing \ac{upf} with a local \ac{dnn} component as shown in~\cref{fig:sysenhance}, which can be configured according to specific applications.

Trivially embedding a \ac{dnn} is unrealistic because \ac{dl} is computationally demanding, and it often requires a complex computing architecture (such as dedicated hardware like GPU and large memory). Although the future \ac{upf} element will be programmable with compute-oriented resources onboard, its computational power shall be still much less than a \ac{cots} server node.
Nevertheless, the size of a \ac{dnn} is usually huge~\cite{Zhang2019DeepLearningMobile}. Hence, a central challenge is how to split a \ac{dnn} into pieces that are network element-suitable. 
Considering this challenge, we propose the following principles when a \ac{dnn} needs to be split and embedded on several \acp{upf} along a forwarding path.
\begin{enumerate}
    \item Principle-1 \emph{proportionally splitting a model}:
    Currently, popular \acp{dnn} easily have millions or higher numbers of \rv{parameters} with dozens of network layers.
    Unlike a server, a \ac{upf} cannot offload the entire model. Splitting a model proportionally can assure that a \ac{upf} will not be overloaded;
    \item Principle-2 \emph{avoiding inter-\ac{upf} traffic explosion}: A \ac{dnn} transforms raw data into high-dimensional feature representations by convolution or other operations. This process usually accompanies temporal variations of data volume transmitted between layers. To avoid the data volume between two \acp{upf} exploding, it is necessary to split a \ac{dnn} at the layers whose temporal data volume is small. \rv{Note that} achieving this objective may contradict to the first one;
    \item Principle-3 \emph{parallelizing local processing and forwarding}: Local \ac{dl} on a \ac{upf} requires data and temporal outputs of the last layer of a \ac{dnn} delivered from the previous \ac{upf}. Caching and waiting for the required inputs take time, especially when the path length increases. Hence, parallelizing the underlying forwarding strategy and efficient inter-\acp{upf} interactions should be considered.    
\end{enumerate}
Bearing with the three principles above, a generic way to split a \ac{dnn} can follow the procedures below: 

The first step is to analyze the parameter size distribution of the \ac{dnn}. Following \rv{Principle-1}, the \ac{dnn} is divided into multiple neural blocks proportional to the capacities of allocated \acp{upf}.

The second step is to adjust the splitting result from the first step and avoid data explosion splitting points located in between two consecutive \acp{upf} as much as possible, \rv{following Principle-2}. For example, the authors in~\cite{Wu21:InNetworkProcessingLowLatency} use the saddle point of filtering rate to determine splitting points where the amount of temporal data is smaller. This tells us which layer should or should not be a splitting point.

The third step is to analyze the computing logic of neural blocks to avoid waiting for other neural blocks' processing, thus wasting \ac{upf}'s computing resources. For example, the low-complexity shortcut path of a bottleneck residual block must wait for data from the high-complexity path. \rv{With following Principle-3}, such processing and forwarding need to be parallelized to avoid caching and waiting. Therefore, the splitting result may be further revised.

\subsection{\rv{\ac{mp}} and \ac{cp} extension}\label{subsec:mcpenhance}
After we introduce the part of \ac{dp} enhancement, we briefly introduce what has to be done at the \ac{mp} and \ac{cp} accordingly for completeness, although this part is not the main focus of this paper.

\subsubsection{\ac{mp} extension}
\ac{mp} has to support orchestrating the new type of \ac{upf} with \ac{dl} capability. Specifically, since a \ac{dl} application usually uses a dedicated \ac{dnn} model, creation and modification of such a \ac{upf} need to involve a service provider, which is not supported in 5G \ac{mp}. In other words, the existing 5G \ac{mp} does not allow to configure \ac{upf} with an additional processing logic for a particular application service. In order to enable this, as highlighted in red in~\cref{fig:sysenhance}, the following minimum extensions are suggested:
\begin{itemize}
 \item \rv{Os-Ma-nfvo}: Over this interface, an application service provider shall be able to publish a set of in-network computing logic for its application services to \rv{\ac{nfv}-\ac{mano} (nfvo)};
 \item \rv{Ve-Vnfm-vnf}: \ac{vnfm} shall be able to receive the request of deploying a \ac{upf} with an in-network computing logic defined by the service provider from \ac{nfv}-\ac{mano}; \ac{vnfm} prepares a \ac{nf} description in terms of the resource availability;
 \item \rv{Nf-Vi}: A \ac{nf} description with the service provider's feature is sent to \ac{vim} who will instantiate a \ac{vnf} instance at the \ac{nfv} infrastructure.
\end{itemize}
After the customized \ac{upf} is instantiated, the instance will register itself or be identified by the relevant management entity (e.g., the element manager of the network) for \emph{configuration, fault, and performance managements}. It is also ready for service provisioning with \ac{cp}. For our case, the extended \ac{mp} deploys the \ac{upf} instances with a split \ac{dnn} for a \ac{dl} application service.

\subsubsection{\ac{cp} extension}
\ac{cp} has to support provision a service that is requested by a \ac{ue} indicating to activate the enhanced feature at the \ac{dp} (i.e., the \ac{upf} with \ac{dl} capability). This requires a series of updated control flows among existing \ac{cp}-\acp{nf}. At least the two aspects below have to be extended:
\begin{itemize}
	\item \emph{Service Discovery}: As shown in~\cref{fig:sysenhance}, this happens between a \ac{ue} and an \ac{easdf}. An \ac{easdf} provides edge application service information that is needed for a \ac{ue}. The information has to be extended to include whether a requested application service supports the in-network computing feature; after receiving the response from \ac{easdf}, a \ac{ue} can decide if it wants to establish a \ac{pdu} session over the enhance \acp{upf};
	\item \emph{Session Management}: As shown in~\cref{fig:sysenhance}, this happens between a \ac{ue} and a \ac{smf} that is responsible for \ac{pdu} session establishment. \ac{smf} has to be extended to support identify, select and configure the new type of \acp{upf} that are enhanced with application-specific processing logic; \ac{smf} shall also react to mobility events of the \ac{ue} for service continuity, \ac{qos} measurement collections and report to other \ac{cp}-\acp{nf} for charging and policy interventions and so on.
\end{itemize}

In summary, \ac{dp} and M-/CP enhancements  together complete the picture of enhancing in-network \ac{dl} capability for \ac{ml} inference application services in 6G.
\begin{figure*}[t!]
    \centerline{\includegraphics{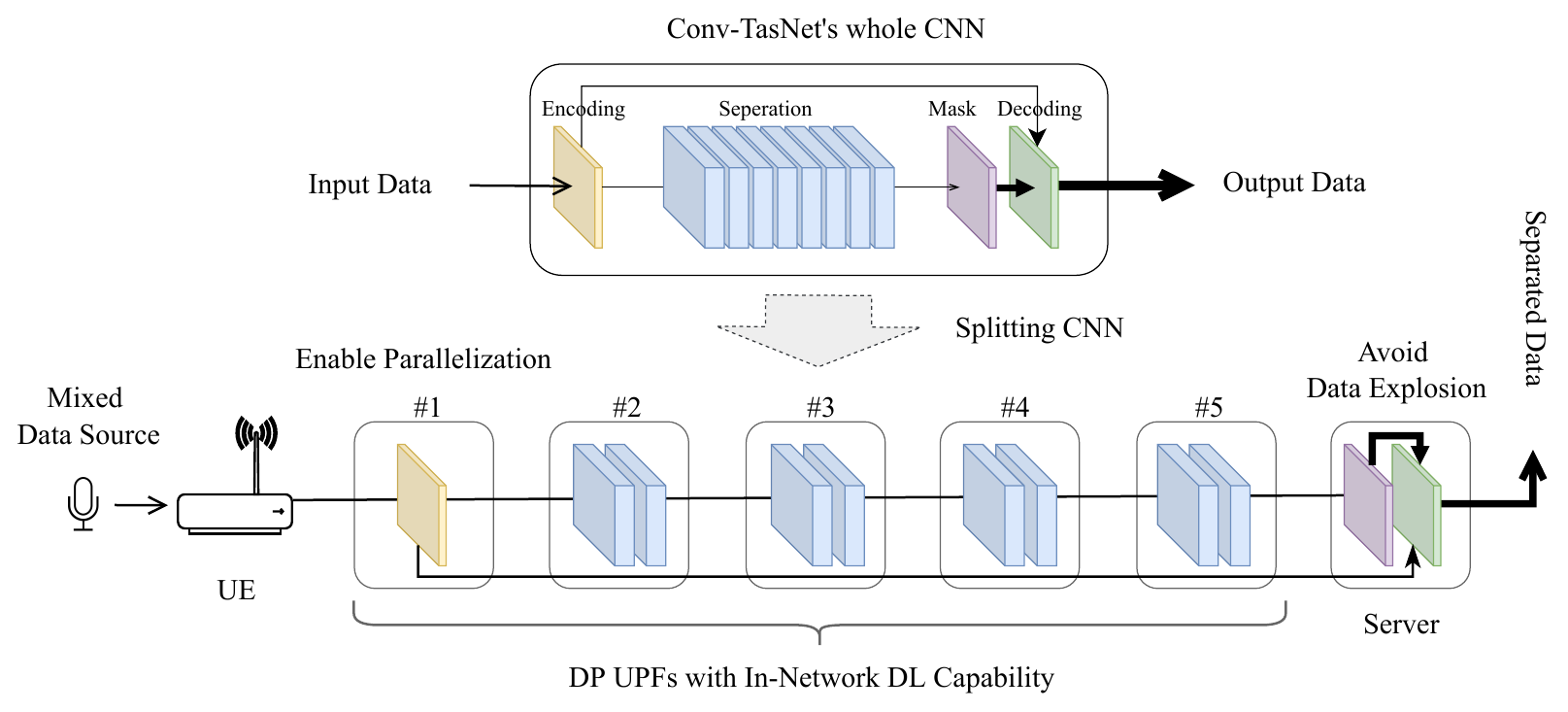}}
	\caption{An illustration of the ultimate splitting of Conv-TasNet. \ac{dp} path length is five, and every \ac{upf} can accommodate maximum of two neural blocks.}
	\label{fig:conv_tasnet_split}
\end{figure*}

\section{A Case Study: \acf{bss} Problem}\label{sec:bss_example}

\subsection{\ac{bss} problem}\label{subsec:bss_intro}
\ac{bss}~\cite{Comon2010HandbookBlindSource} is the process of recovering the source signal from a mixture of observed signals without the aid of information (or with very little information) about the source signals and the mixing process.
Since acoustic signals are naturally interfered with, the observed signals need to be separated to understand each source individually. As introduced, \ac{bss} is a fundamental data pre-processing technique for numerous acoustic applications, such as audio analysis, natural language processing, speech recognition, etc. 
It is also promising if such a pre-processing can be accelerated when collected data is being transmitted via a mobile network before arriving at a server end.

\ac{bss} problem has been studied for years in the field of signal processing. One of the most successful approaches is \ac{ica}~\cite{Comon2010HandbookBlindSource}. Because of its high demand for computational power, \ac{ica} is often executed on a powerful server machine.
Recently, \ac{dl} techniques are adopted to solve the \ac{bss} problem, as an alternative to \ac{ica} methods. 
%
One of the most advanced is a \ac{cnn}-based Conv-TasNet~\cite{Luo2019ConvtasnetSurpassing}, which is recalled upon next.

\subsection{Conv-TasNet preliminary}\label{subsec:intro_TasNet}
Conv-TasNet consists of several convolutional layers as shown in~\cref{fig:conv_tasnet_split}.
The input time-domain signal is encoded into the feature space by the convolutional layer at the beginning (marked with yellow). 
After the first layer, there are eight stacked 1-D convolutional layers as the separation part (marked with light blue) performing the actual data separation jobs.
After the separation part, another convolutional layer (marked with purple) expands the intermediate data to masks whose size will be significantly increased (indicated by a thicker line in~\cref{fig:conv_tasnet_split}).
The last convolutional layer (marked with green) decodes the data into the time-domain and also causes an increase in the length of the data.



Conv-TasNet is a valuable candidate to demonstrate how to accommodate a \ac{dnn} into an in-network scheme. The reasons are as follows. First, Conv-TasNet utilizes lightweight 1-D convolutional blocks to separate the observed signals. Such a lightweight design reduces the overall computational complexity with guaranteed separation accuracy that makes it suitable for deployment in networks. Deriving an in-network solution for it will also be instrumental for other similar \ac{bss} problems. Second, we will see that splitting its \ac{cnn} will face all challenges we mentioned before. Therefore, this exemplifies how the three proposed principles shall be exercised in practice.

\subsection{Splitting Conv-TasNet into \ac{dp}}\label{subsec:splitTasNet}


Refer to the three principles introduced in~\cref{subsec:dpdl}, Conv-TasNet is split into ten neural blocks, as shown in~\cref{fig:conv_tasnet_split}.

Following \rv{Principle-1}, with homogeneous \acp{upf}, we initially split Conv-TasNet into eight even neural blocks, each of which occupies approximately $12.649\%$ out of $0.663$ millions of model parameters defining the entire \ac{cnn}.
Specifically, the first neural block contains the encoding part convolutional layer and a subset of separation part convolution layers; the next six neural blocks correspond to the second to seventh evenly split convolutional layers of the separation part; the eighth neural block contains the rest of the separation part convolutional layers (including mask) and the decoding part. Note that the separation part occupies almost $98.764\%$ parameters while all the other parts share the rest $1.236\%$.

According to \rv{Principle-2}, furthermore, the last neural block (i.e., the rest of the separation part and the mask plus decoding part) enlarges output data size, which leads to traffic explosion to stress network bandwidth to the server node.
Therefore, the decoding part is taken out as a new neural block and merged to the server node. This hides the explosion inside the server node thus avoiding large volume temporal data flooding the network link.

Observing the computing logic of the encoding part, its output is needed by both the separation part and the last decoding part. To prevent the last convolutional layer from waiting for temporal output, \rv{with Principle-3} we split the convolutional layer of the decoding part as a separate neural block, so that its output data can be directly sent to the last in parallel.

Ultimately, from the initial eight neural blocks, with considering \rv{Principle-2 and Principle-3}, the total number of split neural blocks becomes ten. For illustration, \cref{fig:conv_tasnet_split} shows a mapping with its final deployment on \ac{dp} consisting of \acp{upf} and the server node, given that every \ac{upf} can accommodate maximum of two neural blocks, i.e., $\delta$-type is $2$ as how we define later.
\section{Evaluation}\label{sec:evaluation}
\subsection{Experimental setups}
We emulated the proposed enhancements with a network emulator--\ac{comnetsemu} developed in~\cite{XiangOpenSourceTestbed}.
The reason to choose \ac{comnetsemu} is because of its full capability of supporting both \ac{nfv} and \ac{sdn}.
Specifically, we \rv{used} \ac{comnetsemu} to implement the extended \ac{dp} described in~\cref{subsec:dpdl} and \rv{accommodated} the split Conv-TasNet solution introduced~in~\cref{subsec:splitTasNet}.
The virtualized network node created by \ac{comnetsemu} is equivalent to a \ac{upf} with a split \ac{dnn} component.
Hence, a series of virtualized nodes form the \ac{dp} running the data separation service.
Additionally, the \ac{sdn} controller mimics the role of extended \ac{mp} and \ac{cp} introduced in~\cref{subsec:mcpenhance}.

Data \rv{was} transmitted with \ac{udp} and the bandwidth between \acp{upf} \rv{was} $1$~\textit{Gbps} with propagation delay $10$~\textit{ms}. 
These are normal conditions in an average network system.
Since we mainly focused on the split of \rv{\ac{nn}} models, we \rv{assumed} that possible transmission failures \rv{were} handled by lower-layer protocols.
All emulations were done on a \ac{cots} server with an Intel(R) Xeon(R) Gold 6148 CPU 2.40GHz and 4GB RAM using Ubuntu 18.04.4 LTS.

\subsubsection{Scenarios}\label{subsec:evaluation_scenarios}

\begin{table}[t!]
    \def\arraystretch{1.2}
    \centering
    \caption{Distributions of ten neural block$\sharp$ on network elements within two network path lengths. The first neural block always occupies one \ac{upf} and the last neural block is always on server node.}\label{tab:evaluation_scenario}
    \begin{tabular}{C{0.7cm}|C{1.9cm}|C{2.2cm}|C{2.2cm}}
    \toprule[1pt]
    $\delta$-Type  & Max. Neural Block$\sharp$ per \ac{upf} & Neural Block$\sharp$ on Server (\ac{upf}$\sharp=3$) & Neural Block$\sharp$ on Server  (\ac{upf}$\sharp=5$) \\
    \midrule[0.5pt]
    0 & NULL & 10  & 10  \\
    \hline
    1 & 1  & 7  & 5  \\
    \hline
    2 & 2  & 5   &  1\\
    \hline
    3 & 3  & 3   &  1\\
    \hline
    4 & 4  & 1 & 1 \\
    \bottomrule[1pt]
    \end{tabular}
\end{table}

We \rv{evaluated} under a linear topology with two different \ac{dp} path lengths: three and five hops (i.e., namely \emph{3H-N} and \emph{5H-N}), respectively.
This is in the typical range of the \ac{upf} numbers in an operator's network from a \ac{ue} to data networks.
We \rv{considered} five types of \acp{upf}, numbered with an integer value $\delta$ from $0\rightarrow4$. 
Each $\delta$-type represents a \ac{upf} with the capacity to accommodate maximally $\delta$ numbers of Conv-TasNet's neural blocks onboard.
In particular, the scenario of 0-type \acp{upf} \rv{was} used as the baseline system, since $\delta$ equals zero means that the network has no in-network \ac{dl} enhancement, in this case, it is equivalent to the original Conv-TasNet scheme in~\cite{Luo2019ConvtasnetSurpassing}.
\cref{tab:evaluation_scenario} gives the detailed network topology configurations consisting of different $\delta$-type \acp{upf} and a server end.
In addition to comparing with Conv-TasNet without in-network \ac{dl} capability, we also \rv{selected} a recent non-\ac{dl}-based solution -- \rv{\ac{pica}~\cite{picaextention}} for comparison.

For each configuration in~\cref{tab:evaluation_scenario}, we performed 50 tests i.e., 50 randomly selected datasets will be processed with our implementation in the \ac{comnetsemu} emulator.

\subsubsection{Metrics}\label{subsec:evaluation_metrics}
The first metric is \textit{residual time}, which tells the \emph{residual} time taken by the last node (i.e. server node) to finalize the entire data separation jobs. The smaller the residual time, the more jobs were done by the intermediate network nodes. Therefore, it measures the acceleration speed to the \ac{dl} service application service.

The second metric is \textit{\ac{sdr}}, which is used to evaluate the separation accuracy of Conv-TasNet service.
The \ac{sdr} definition is the most widely used metric nowadays because different types of errors, i.e., interference, noise, and artifacts errors, are comprehensively considered.
This metric examines whether the accuracy of neural networks has been affected by in-network computing.

\subsubsection{Dataset}\label{subsec:evaluation_dataset}
We \rv{picked} a public open dataset from~\cite{Purohit_DCASE2019_01}, called \ac{mimii}. 
The main reasons for choosing the \ac{mimii} dataset are as follows.
First, it is a well-known dataset that is widely referenced in research and engineering for acoustic machine anomaly detection.
In addition, it exactly reflects the theme of this work, i.e., processing the most relevant acoustic data of factory machines from industry rather than random acoustic data such as music or daily life data.

This dataset has collected $26092$ normal and anomalous operating sound data of four types of machines.
$2$-second-long audio of each segment is used, which is single-channel with a sample rate of $16$kHz. The size of one data source is $32k$.


\subsection{Residual computing time and separation quality}

\begin{figure}[t]
	\centerline{\includegraphics[width=\columnwidth]{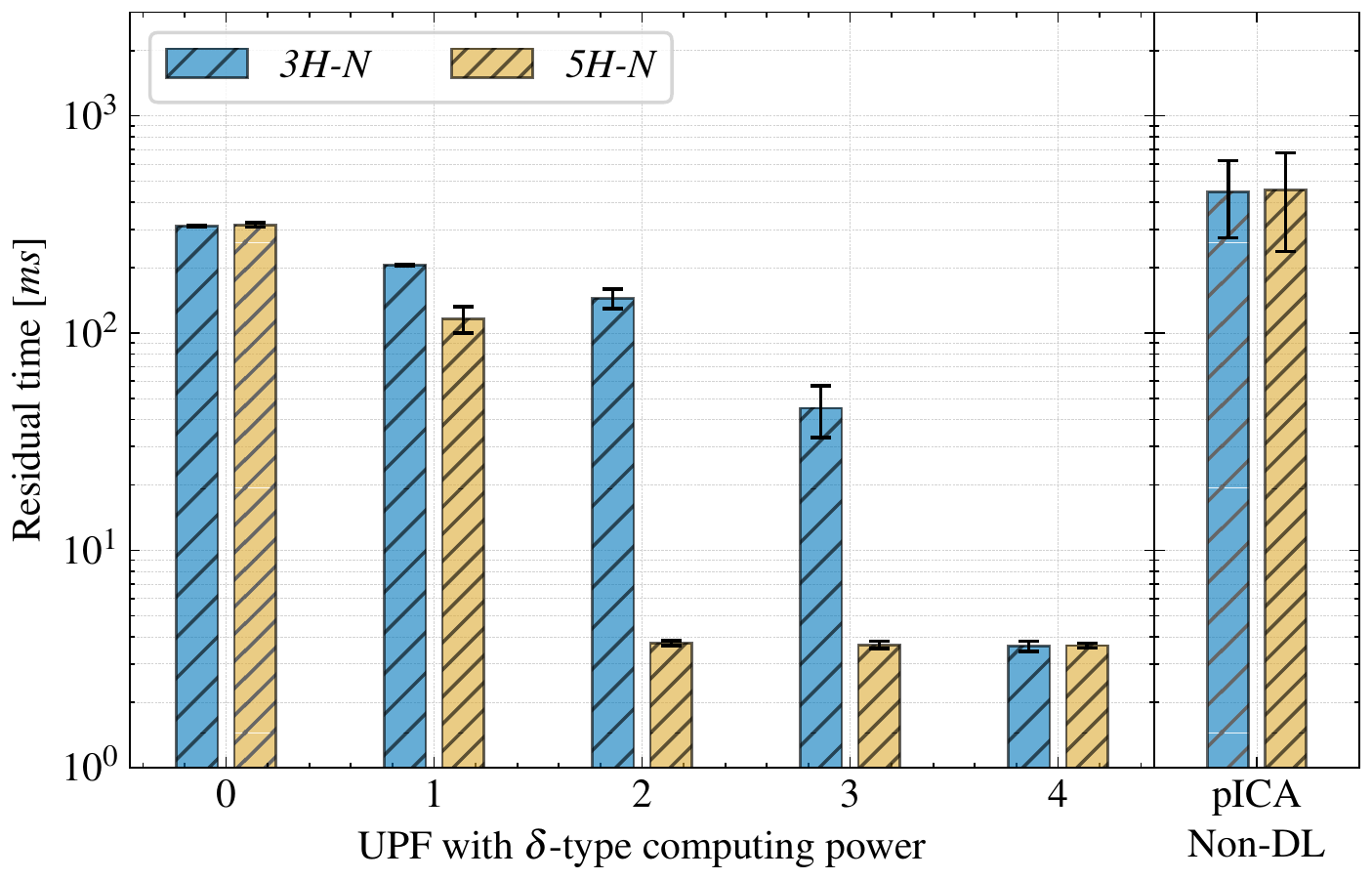}}
	\caption{Residual time of the \ac{dl}-based Conv-TasNet and the non-\ac{dl}-based \ac{pica} on \ac{dp}.}
	\label{fig:residule_time_bar}
\end{figure}

\cref{fig:residule_time_bar} shows the residual time on the server node when using the \ac{dl}-based Conv-TasNet (with and without the proposed \ac{dp} enhancement) and \rv{\ac{pica}}, which is an \ac{ica} algorithm but not based on \ac{dl}, under the two \ac{dp} lengths with different $\delta$-type \acp{upf}.

The first observation is that no matter if using \ac{dp} enhancement, Conv-TasNet \rv{showed} smaller residual time than \ac{pica} has, e.g., $310~ms$ of $0$-type \ac{upf} v.s. $447~ms$ of \ac{pica} in \emph{3H-N}. 
This is because of the nature of the Conv-TasNet method, where Conv-TasNet \rv{needed} less input data thus less overall workloads. This justifies that Conv-TasNet is representative to the benefits of \ac{dl}-based methods.

Secondly, boosted with the proposed \ac{dp} enhancement, when the computing resources gradually \rv{increased}, the residual time on the server node \rv{could} be further reduced. Specifically, given the same \ac{dp} path length, the more powerful the \acp{upf} \rv{were} used, the smaller the residual time \rv{would} be spent on the server node.
The residual time was reduced by $98.83\%$ when the $\delta$-type went from zero to four in the \emph{3H-N} scenario.
Similarly, given the same $\delta$-type \acp{upf}, the longer the \ac{dp} path length, the smaller the residual time \rv{would} be spent on the server node. 
For instance, by increasing the path length of $2$-type \acp{upf} from three to five, the residual time fell from $144.34~ms$ to $3.74~ms$.

Moreover, it \rv{showed} an upper bound of the acceleration, influenced by both the computing resources of the \ac{upf} and the offloaded neural blocks.
Specifically, in \emph{5H-N} case, the residual time \rv{reduced} to $3~ms$ and \rv{did} not change anymore since \ac{upf} type \rv{was} two and onward. 
This is because according to the Conv-TasNet splitting in~\cref{subsec:splitTasNet}, maximally only nine intermediate nodes \rv{were} needed to accommodate all split neural blocks even if the \ac{upf} $\delta$-type equals one. 
Hence, adding more intermediate \acp{upf} will not further improve the acceleration (i.e., reducing residual time on the server node).

\begin{figure}[t]
	\centerline{\includegraphics[width=\columnwidth]{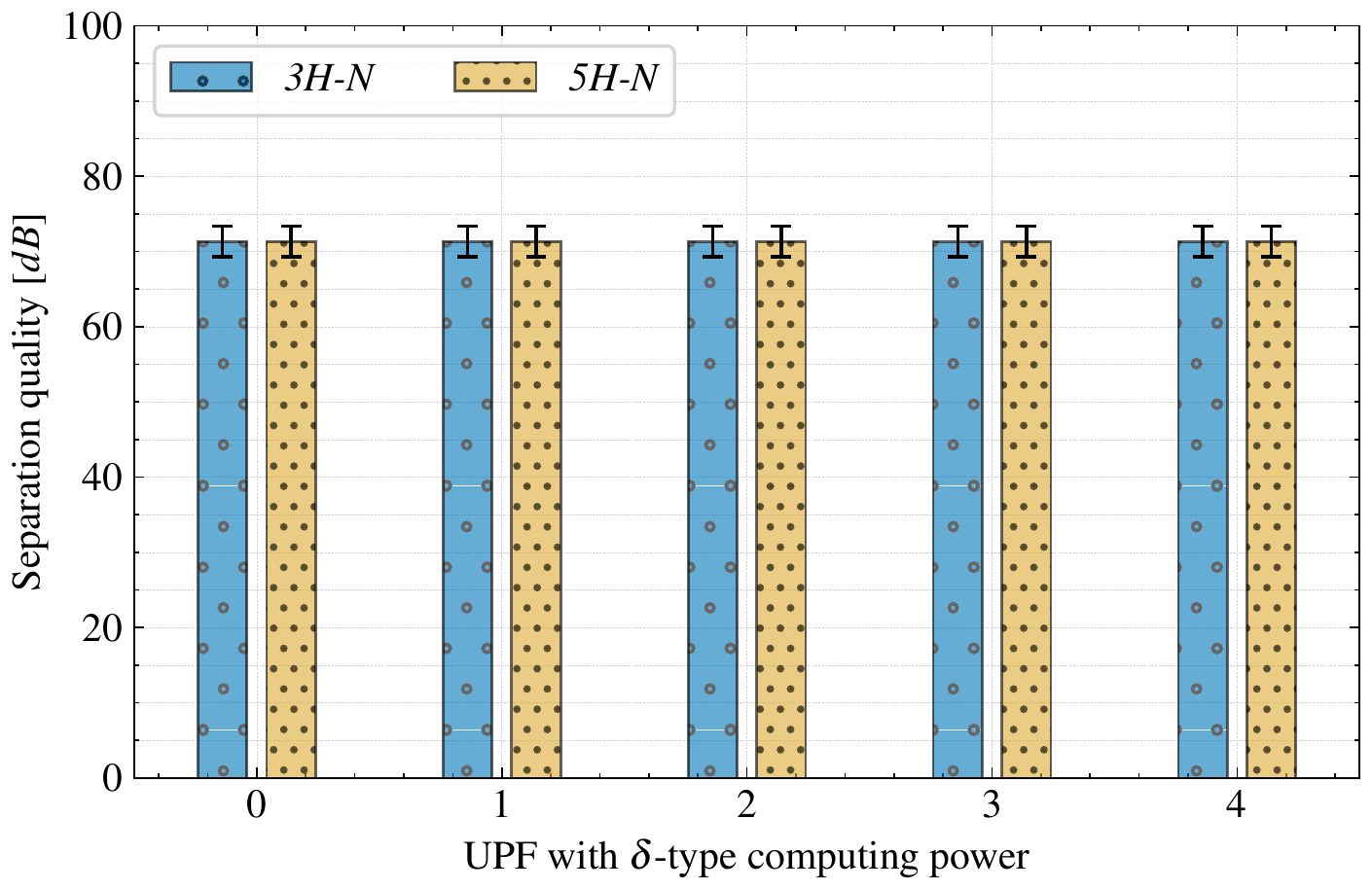}}
	\caption{The separation quality of DL-based ConvTasNet on DP.}
	\label{fig:sdr_bar}
\end{figure}

Last but not least, \cref{fig:sdr_bar} shows the achievable separation quality, \ac{sdr}, with the proposed in-network scheme within the two different \ac{dp} path lengths. 
All \acp{sdr} \rv{had} no noticeable loss (about $71~dB$) compared with the original centralized Conv-TasNet (i.e., the \ac{upf} with $\delta$-type is zero).
This is because although we split the whole \ac{cnn} into several neural blocks, its structure across all \acp{upf} \rv{was} not modified.
This suggests that running the \ac{dl} inference jobs with the \ac{dp} enhancement can be a promising potential scheme.

\subsection{Traffic explosion avoidance}
\begin{figure}[t]
	\centerline{\includegraphics[width=\columnwidth]{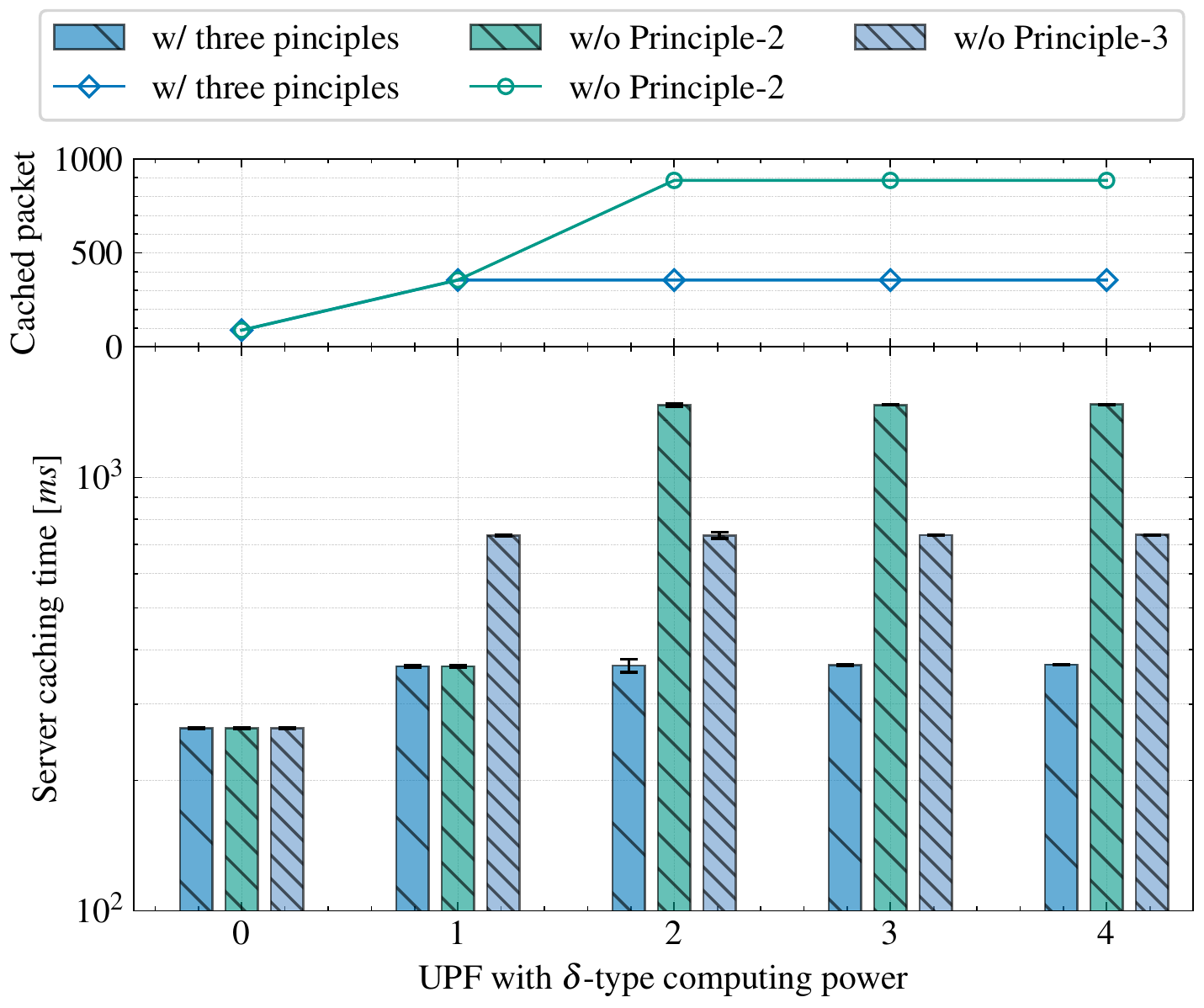}}
	\caption{The caching time and the number of cached packets on the server in \emph{5H-N}. 
	}
	\label{fig:cache_time_bar}
\end{figure}

Between two adjacent \acp{upf}, since the neural blocks on a \ac{upf} may increase the size of output data, this may cause traffic explosion on the link to the next \ac{upf}.
The traffic explosion can be mitigated by carefully splitting the \ac{cnn} at the layers whose temporal data size is smaller, i.e., Principle-2 \emph{avoiding inter-\ac{upf} traffic explosion} introduced in~\cref{subsec:dpdl}.

As shown in~\cref{fig:cache_time_bar}, without following Principle-2, the number of packets cached by the server \rv{grew} from $356$ to $886$, and the caching time \rv{increased} four times from $366~ms$ to $1472~ms$ when $\delta$ \rv{was} bigger than one. 
As we explained, this is because the mask part (the purple block in~\cref{fig:conv_tasnet_split}) is a splitting point that will temporally increase the output data size. 
Hence, we merged this layer to the server end following Principle-2, this \rv{avoided} sending large data traffic to the network link. 
As we can see, the caching time \rv{remained} $366~ms$, which \rv{was} slightly higher than $263~ms$ of 0-type \ac{upf} only. 
This again confirms the influence of the splitting points to the traffic generated in the network.

\subsection{Processing and forwarding parallelization}

We also measured the caching time at the server waiting for caching required processing data, presented in~\cref{fig:cache_time_bar}.
This caching time can be improved by Principle-3 in~\cref{subsec:dpdl}.

\cref{fig:cache_time_bar} provides a comparison on processing and forwarding with and without Principle-3 in \emph{5H-N} scenario.
With the applied parallelization scheme, the server \rv{started} receiving the encoding data at the same time as the \ac{upf}, therefore, the idle time \rv{was} only about $103~ms$, a $78.09\%$ reduction.
In contrast, without parallelization, the caching time of the server \rv{increaseed} from $263~ms$ to $733~ms$, i.e., the server was idle approximately $470~ms$.
This is because the encoding part and the first separator belong to the same neural block, which causes \ac{upf} to forward the intermediate data to the server only after it completes all operations. 
This confirms the necessity of such a parallelization consideration and the result confirms its effectiveness to improve the efficiency at the server end.

\section{Conclusion}\label{sec:conclusion}

This paper presents a \ac{dp} enhancement with accommodating an application-specific in-network \acrlong{dl}.
In particular, the focus is on how to split a trained \ac{dnn} model so that it can fit into \acp{upf} along a forwarding path.
Taking a specific use case -- \acrlong{bss} -- as an example, we demonstrated how a \ac{cnn} used in an existing work shall be split with our proposed principles. 
Furthermore, we implemented the suggested scheme with a full-stack emulator -- \acrfull{comnetsemu} and showed that the proposed enhancement indeed yields noticeable caching accelerations and reduces the workload left to the server end.
Since the structure of the \ac{cnn} is untouched, we believe that the suggested strategy/principles shall be applicable to other \ac{dl} application services based on \ac{dnn}. 
As an initial trial, our work also reveals further challenges that require deeper investigations from the whole community when considering splitting a \ac{cnn} onto in-network elements, to support \ac{ai} application in future 6G.
\section*{Acknowledgment}
This work has been funded by the Federal Ministry of Education and Research of Germany (Software Campus Net-BliSS 01IS17044), and by the European Commission through the H2020 projects Hexa-X (Grant Agreement no. 101015956).
It has also been partially funded by the German Research Foundation (DFG, Deutsche Forschungsgemeinschaft) as part of Germany’s Excellence Strategy – EXC 2050/1 – Project ID 390696704 – Cluster of Excellence “Centre for Tactile Internet with Human-in-the-Loop” (CeTI) of Technische Universität Dresden.

\bibliographystyle{IEEEtran}
\bibliography{bibtex}

\begin{thebibliography}{10}
\providecommand{\url}[1]{#1}
\csname url@samestyle\endcsname
\providecommand{\newblock}{\relax}
\providecommand{\bibinfo}[2]{#2}
\providecommand{\BIBentrySTDinterwordspacing}{\spaceskip=0pt\relax}
\providecommand{\BIBentryALTinterwordstretchfactor}{4}
\providecommand{\BIBentryALTinterwordspacing}{\spaceskip=\fontdimen2\font plus
\BIBentryALTinterwordstretchfactor\fontdimen3\font minus
  \fontdimen4\font\relax}
\providecommand{\BIBforeignlanguage}[2]{{%
\expandafter\ifx\csname l@#1\endcsname\relax
\typeout{** WARNING: IEEEtran.bst: No hyphenation pattern has been}%
\typeout{** loaded for the language `#1'. Using the pattern for}%
\typeout{** the default language instead.}%
\else
\language=\csname l@#1\endcsname
\fi
#2}}
\providecommand{\BIBdecl}{\relax}
\BIBdecl

\bibitem{Lao2021ATPnetworkAggregation}
C.~Lao, Y.~Le, K.~Mahajan, Y.~Chen, W.~Wu, A.~Akella, and M.~Swift, ``{ATP}:
  In-network aggregation for multi-tenant learning,'' in \emph{18th USENIX
  Symposium on Networked Systems Design and Implementation (NSDI 21)}.\hskip
  1em plus 0.5em minus 0.4em\relax USENIX Association, Apr. 2021, pp. 741--761.

\bibitem{Dang2020P4xosConsensusas}
H.~T. Dang, P.~Bressana, H.~Wang, K.~S. Lee, N.~Zilberman, H.~Weatherspoon,
  M.~Canini, F.~Pedone, and R.~Soul{\'e}, ``P4xos: Consensus as a network
  service,'' \emph{IEEE/ACM Transactions on Networking}, vol.~28, no.~4, pp.
  1726--1738, 2020.

\bibitem{Luo2019ConvtasnetSurpassing}
Y.~Luo and N.~Mesgarani, ``Conv-tasnet: Surpassing ideal time--frequency
  magnitude masking for speech separation,'' \emph{IEEE/ACM transactions on
  audio, speech, and language processing}, vol.~27, no.~8, pp. 1256--1266,
  2019.

\bibitem{picaextention}
H.~Wu, Y.~Shen, X.~Xiao, G.~T. Nguyen, A.~Hecker, and F.~H.~P. Fitzek,
  ``Accelerating industrial iot acoustic data separation with in-network
  computing,'' \emph{IEEE Internet of Things Journal}, pp. 1--16, 2022.

\bibitem{Wu2021ComputingMeetsNetwork}
H.~Wu, Z.~Xiang, G.~T. Nguyen, Y.~Shen, and F.~H. Fitzek, ``Computing meets
  network: Coin-aware offloading for data-intensive blind source separation,''
  \emph{IEEE Network}, vol.~35, no.~5, pp. 21--27, 2021.

\bibitem{Xu2021GRACECompressedCommunication}
H.~Xu, C.-Y. Ho, A.~M. Abdelmoniem, A.~Dutta, E.~H. Bergou, K.~Karatsenidis,
  M.~Canini, and P.~Kalnis, ``Grace: A compressed communication framework for
  distributed machine learning,'' in \emph{2021 IEEE 41st International
  Conference on Distributed Computing Systems (ICDCS)}, 2021, pp. 561--572.

\bibitem{DBLP:journals/corr/abs-2107-03428}
\BIBentryALTinterwordspacing
S.~Trindade, L.~F. Bittencourt, and N.~L.~S. da~Fonseca, ``Management of
  resource at the network edge for federated learning,'' \emph{CoRR}, vol.
  abs/2107.03428, 2021. [Online]. Available:
  \url{https://arxiv.org/abs/2107.03428}
\BIBentrySTDinterwordspacing

\bibitem{Liang2020edgebaseddeep}
F.~Liang, W.~Yu, X.~Liu, D.~Griffith, and N.~Golmie, ``Toward edge-based deep
  learning in industrial internet of things,'' \emph{IEEE Internet of Things
  Journal}, vol.~7, no.~5, pp. 4329--4341, 2020.

\bibitem{Xavier2021ProgrammableSwitchesNetworking}
B.~M. Xavier, R.~S. Guimarães, G.~Comarela, and M.~Martinello, ``Programmable
  switches for in-networking classification,'' in \emph{IEEE INFOCOM 2021 -
  IEEE Conference on Computer Communications}, 2021, pp. 1--10.

\bibitem{Saquetti2021NetworkIntelligenceRunning}
M.~Saquetti, R.~Canofre, A.~F. Lorenzon, F.~D. Rossi, J.~R. Azambuja,
  W.~Cordeiro, and M.~C. Luizelli, ``Toward in-network intelligence: Running
  distributed artificial neural networks in the data plane,'' \emph{IEEE
  Communications Letters}, vol.~25, no.~11, pp. 3551--3555, 2021.

\bibitem{Zhang2019DeepLearningMobile}
C.~Zhang, P.~Patras, and H.~Haddadi, ``Deep learning in mobile and wireless
  networking: A survey,'' \emph{IEEE Communications surveys \& tutorials},
  vol.~21, no.~3, pp. 2224--2287, 2019.

\bibitem{Wu21:InNetworkProcessingLowLatency}
H.~Wu, J.~He, M.~Tömösközi, Z.~Xiang, and F.~H. Fitzek, ``In-network
  processing for low-latency industrial anomaly detection in softwarized
  networks,'' in \emph{2021 IEEE Global Communications Conference (GLOBECOM)},
  2021, pp. 01--07.

\bibitem{Comon2010HandbookBlindSource}
P.~Comon and C.~Jutten, \emph{Handbook of Blind Source Separation: Independent
  component analysis and applications}.\hskip 1em plus 0.5em minus 0.4em\relax
  Academic press, 2010.

\bibitem{XiangOpenSourceTestbed}
Z.~Xiang, S.~Pandi, J.~Cabrera, F.~Granelli, P.~Seeling, and F.~H.~P. Fitzek,
  ``An open source testbed for virtualized communication networks,'' \emph{IEEE
  Communications Magazine}, vol.~59, no.~2, pp. 77--83, 2021.

\bibitem{Purohit_DCASE2019_01}
H.~Purohit, R.~Tanabe, T.~Ichige, T.~Endo, Y.~Nikaido, K.~Suefusa, and
  Y.~Kawaguchi, ``{MIMII Dataset}: Sound dataset for malfunctioning industrial
  machine investigation and inspection,'' in \emph{Proceedings of the Detection
  and Classification of Acoustic Scenes and Events 2019 Workshop
  ({DCASE2019})}, 11 2019, pp. 209--213.

\end{thebibliography}

\vskip 0pt plus -1fil
\begin{IEEEbiographynophoto}
{Jia He} 
is a graduate student with the Deutsche Telekom Chair of Communication Networks at Technische Universität Dresden (TU Dresden), Germany, since 2017.
He received his Bachelor of Science degree from Xidian University, China, in 2015.
His interests during his studies are data analysis in networks.
\end{IEEEbiographynophoto}

\vskip -2\baselineskip plus -1fil
\begin{IEEEbiographynophoto}
{Huanzhuo Wu} 
is a research fellow with the Deutsche Telekom Chair of Communication Networks at Technische Universität Dresden (TU Dresden), Germany.
He received his M. Sc. degree in Computer Science at TU Dresden, in 2016.
His research interest is data analysis on Blind Source Separation and in-network computing.
\end{IEEEbiographynophoto}

\vskip -2\baselineskip plus -1fil
\begin{IEEEbiographynophoto}
{Xun Xiao} 
is now a senior researcher with Munich Research Center of Huawei Technologies, Germany. He received his Bachelor's and Master's degrees in Computer Science both in South-Central University for Nationalities in 2006 and 2009, respectively, and his Ph.D. degree in Computer Science from the City University of Hong Kong in 2012. From 2012 to 2014, he was a postdoc researcher at Max Planck Institute of Molecular Cell Biology and Genetics (MPI-CBG). His research interests mainly focus on algorithms, networking, and computing in distributed systems.
\end{IEEEbiographynophoto}

\vskip -2\baselineskip plus -1fil
\begin{IEEEbiographynophoto}
{Riccardo Bassoli} 
is a \rv{junior professor} with the Deutsche Telekom Chair of Communication Networks \rv{and the head of the Quantum Communication Networks Research Group} at Technische Universität Dresden (TU Dresden), Germany. He received his Ph.D. degree from 5G Innovation Centre (5GIC) at the University of Surrey (UK), in 2016.
\end{IEEEbiographynophoto}


\vskip -2\baselineskip plus -1fil
\begin{IEEEbiographynophoto}
{Frank H. P. Fitzek} 
is a professor and head of the Deutsche Telekom Chair of Communication Networks at TU Dresden, Germany.
He received his Dipl.-Ing. degree in electrical engineering from the Aachen University of Technology RWTH, Germany, in 1997 and his Dr.-Ing. degree in electrical engineering from Technical University Berlin.
\end{IEEEbiographynophoto}


\end{document}